\documentclass[%
prd,%
preprint%
,secnumarabic%
,amssymb, amsmath,nobibnotes, aps]{revtex4}
\usepackage{epsfig}%
\usepackage{graphicx}%
\expandafter\ifx\csname package@font\endcsname\relax\else
 \expandafter\expandafter
 \expandafter\usepackage
 \expandafter\expandafter
 \expandafter{\csname package@font\endcsname}%
\fi

\begin{document}

\title{ Dark energy from Neutrinos and Standard Model Higgs potential}%

\author{  Gaetano Lambiase $^a$ , Hiranmaya Mishra $^b$ and  Subhendra Mohanty $^b$}%
%\email{}
\affiliation{$^a$ University of Salerno, Baronisi, Italy,\\
\& INFN , Sezione di Napoli, Italy \\$^b$ Physical Research Laboratory, Ahmedabad 380009,
India.\\
}
%\date{August 2001}%
\def\be{\begin{equation}}
\def\ee{\end{equation}}
\def\al{\alpha}
\def\bea{\begin{eqnarray}}
\def\eea{\end{eqnarray}}

\begin{abstract}
If neutrino mass is a function of the Higgs potential then minimum of the total thermodynamic potential  $\Omega$ (which is the Higgs potential minus the neutrino pressure) can shift from the standard electro-weak vev $v=246.2$ GeV by a small amount which depends on the neutrino pressure. If the neutrino mass is a very steep function of the Higgs field then the equilibrium thermodynamic potential can act like the dark energy with $\omega \simeq -1$.
Choosing  the neutrino mass as logarithmic function of the Higgs field and a heavy mass scale, we find that the correct magnitude of the cosmological density of the present universe $\rho_\lambda \simeq (0.002 eV)^4$ is obtained by choosing the heavy mass at the GUT scale.

\end{abstract}

\maketitle
\section{Introduction}
The numerical coincidence in the cosmological dark energy density $\rho_\lambda=(0.002 {\rm eV})^4$ and the neutrino mass scale $\Delta m^2=2.5\times  10^{-3} {\rm eV}^2$ which solves  the atmospheric neutrino puzzle has led to the ideas that the minimum of the dark energy potential is set by the dark energy dependent neutrino mass \cite{Fardon, Peccei,Bi, Barger,Cirelli, Brookfield,Afshordi,Takahashi, Fardon2,Brookfield2, Kaplinghat,Ma,Bjaelde, Bernadini, Bhatt}. There is a numerical relation between the GUT scale  $M=10^{16} {\rm GeV}$  the weak scale $v=246.1 {\rm GeV}$ and the dark energy density $\rho_\lambda \sim v^8/M^4$. This motivates us to use relate the Higgs potential to the dark energy. Since neutrinos are the most abundant particles with mass, it is natural to use the neutrino-Higgs interaction to shift the  Higgs field from its  minimum of the  potential $V_\phi=(\lambda/4)(\phi^2-v^2) $. The pressure of the neutrino fluid depends on the neutrino mass. If the neutrino mass is a function of the Higgs field then the field rolls down to the lowest point in the total thermodynamic potential i.e the Higgs field dynamically adjusts to a value which maximizes the total pressure of the Higgs-neutrino fluid. Maximizing the total pressure results in the thermodynamic potential minima shifted from $\langle \phi \rangle =v$ to $\langle \phi \rangle =v + \sigma_m$. For this scenario to work and give the correct dark energy density the neutrino mass as function of the Higgs field must have a very steep gradient. We have considered logarithmic neutrino mass functions which meet this criterion. The neutrino mass function has a mass scale M, and dimensionless small parameter $\epsilon$ as free parameters. To give the correct dark energy density it turns out that $M$ must be close to the  GUT scale owing to the numerical relation $(\rho_\lambda)^{1/4} \sim v^2/M$. Neutrino mass in this scenario turns out to be $m_\nu  =2\times 10^{-3}{\rm eV}$ if neutrinos are assumed to be non-relativistic. This is the  range consistent with the solution of the atmospheric neutrino problem. One  well known problem of the neutrino couplings with low mass quintessence type fields is that they give rise to a long range attractive force between neutrinos which could lead to an instability as pointed out in reference \cite{Afshordi}. In  our scenario the Higgs coupling will give a contact interaction between neutrinos which does not lead to formation of neutrino nuggets and clustering at astrophysical scales.

\section{Thermodynamic potential of Higgs-neutrino system}

The Higgs potential is assumed to be at zero temperature and the thermodynamic potential  equilibrium between the Higgs and the neutrino pressures is due to the Higgs dependence of the neutrino mass which affects the neutrino pressure.  In the absence of the neutrino coupling the Higgs field attains a vev $ \langle \Phi \rangle =v/\sqrt{2}$ such that the Higgs pressure $P_\phi=0$,

The Higgs field $\Phi$ is assumed to be at zero temperature and its thermodynamic potential, in the unitary gauge
where $\Phi=\frac{1}{\sqrt{2}}(0,v+\phi)^T$, is
\be
\Omega_\phi=-P_\phi=\frac{\lambda}{4}( \phi ^2-v^2)^2
\ee
where $v=246.2$ Gev is the scale of electroweak symmetry breaking.
 When the neutrino mass is a function of the Higgs field then the net thermodynamic
 potential
 \be
  \Omega\equiv \Omega_\nu(\phi) + \Omega_\phi(\phi)
\ee
   can have a minima at a value of $\phi_m \neq v$.
 If $m(\phi)$ is of such a form that at the minima  the thermodynamic potential is positive,
 $\Omega(\phi_m)>0$ then we have an effective negative pressure from the Higgs-neutrino interaction.
  The thermodynamic potential has a minima at $\phi=\phi_m$ where
\be
\Omega(\phi_m)^\prime= \Omega_\phi(\phi_m)^\prime + \frac{\partial\Omega_\nu}{\partial m}\,m(\phi_m)^\prime=0
\label{Omegaprime}
\ee
where prime denotes derivative w.r.t $\phi$.
We look for a  functions $m(\phi)$ for which the solution $\phi_m$ of equation (\ref{Omegaprime}) is such that
\be
\Omega_\phi(\phi_m) + \Omega_\nu(\phi_m) >0
\ee
We expand the Higgs field close to the minima of $\Omega_\phi$, by taking $\phi=v +\sigma$.
If the minima occurs at a non-zero $\sigma_m$ then we know that $\Omega_\phi(\sigma_m) > 0$ (from the form of the Higgs potential). We also know that the pressure of the neutrino fluid is always positive which means that $\Omega_\nu(\sigma_m)=-P_\nu <0$.  Moreover the magnitude of the  pressure  $P_\nu$ of relativistic neutrinos  is five orders of magnitude smaller than the dark energy density $\rho_\lambda=(0.002 \rm{eV})^4$ (and  even smaller for non-relativistic neutrinos). This implies that at the minima of $\Omega$ the Higgs potential must dominate over the neutrino potential $\Omega_\phi(\sigma_m)=\lambda v^2 \sigma_m^2 \gg \Omega_\nu(\sigma_m)$ therefore
\be
\rho_\lambda = \lambda v^2 \sigma_m^2
\label{rho}
\ee
implying that the Higgs minima should be shifted from $\phi=v$ by the amount
\be
\sigma_m= \pm\, \frac{1}{\sqrt{\lambda}}\,  1.6 \times 10^{-17} {\rm eV}
\ee
in order to explain the observed cosmological acceleration. The neutrino mass function $m(\phi)$ must be a steep function of the the Higgs field, $m(\sigma_m)^\prime \gg 1$ in order to satisfy the simultaneous requirements $\Omega_\phi(\sigma_m) \gg -\Omega_\nu(\sigma_m)$ and $\Omega_\phi(\sigma_m)^\prime+\Omega_\nu(\sigma_m)^\prime=0$.

The effective equation of state of the Higgs-neutrino fluid is
\be
\omega=\frac{P_\nu(\sigma_m) -V_\phi(\sigma_m)}{\rho_\nu(\sigma_m)+V_\phi(\sigma_m)} \simeq -1
\label{eos}
\ee
as $P_\nu, \rho_\nu \ll V_\phi(\sigma_m)$. The Higgs mass $m_H^2=V_\phi^{\prime \prime}(\sigma_m)\simeq 4 \lambda v^2$ is much larger than the Hubble rate and the Higgs field stays at the bottom of the potential as the universe expands and the value of the minima $\sigma_m$ changes with cosmic time. The change in the pressure of the Higgs-neutrino fluid with time is related to the change in the $\sigma_m$ with time as
\be
\frac{d P}{d t}= \left( \frac{\partial }{\partial \sigma}(-V_\phi(\sigma) + P_\nu(\sigma))\right) \frac{\partial \sigma}{\partial T} \frac{\partial T}{\partial t}
\label{dp}
\ee
 The first bracket in the {\it rhs}  of equation (\ref{dp}) dynamically adjusts to zero at thermodynamic equilibrium. The parameters in the potential must allow the solution of $(-V_\phi + P_\nu)^\prime=0$ in  the present era. This explains the 'why now'  question about why the dark energy dominates in the present era. Once the pressure dynamically adjusts to be stationary the density also is stationary owing as the equation of state $\omega\simeq -1$,
 \be
 \frac{d \rho}{dt} \simeq - \frac{dP}{dt}=0
 \ee
 We note that the thermodynamic principle we apply is that the system adjusts to maximize the pressure (or minimize the thermodynamic potential ). A justification of this from first principles can be found in references \cite{Kapusta,Landau,Reif}. We assume that the Higgs and neutrino pressures are coupled through the neutrino mass and the Higgs field in the potential adjusts such as to be at the minima of the thermodynamic potential $\Omega= V_\phi- P_\nu$. This principle of minimizing the thermodynamic potential is also advocated for the case of mass varying neutrinos in reference \cite{Chitov}. The thermodynamics of stable structures formed from mass varying particles is discussed in \cite{Bertolami}.

Neutrinos decouple from thermal equilibrium with the radiation fluid at a temperature $T=1 $MeV
when their distribution function is the ultra-relativistic Fermi-Dirac distribution, $f(p)= (\exp(p/T)+1)^{-1}$. Below the temperature of 1 MeV both temperature and momenta red-shift with the scale factor as $1/a$ and the distribution retains the same form even though they at the present temperature they may be non-relativistic.
The thermodynamic potential of neutrinos which decoupled when they were ultra-relativistic can be written as
\be
\Omega_\nu(\phi)=-P_\nu = -g_i \int \frac{d^3 p}{(2 \pi)^3} \,\frac{1}{3}\left(\frac{p^2}{E}\right) f(p)
%=-2 \int \frac{d^3p}{(2 \pi)^3}\left[\epsilon+ \frac{1}{\beta} \ln\left( 1 + e^{-\beta \epsilon}\right)\right]
\label{Omeganu}
\ee
where $E=\sqrt{m(\phi)^2 +  p^2}$. Here $f(p)$ is the statistical distribution function of neutrinos. If neutrinos have only weak interactions with other particles and decouple from thermal equilibrium at $T \sim 1 {\rm MeV}$ then  their distribution function remains relativistic $f(p)=(\exp(p/T) +1)^{-1}$  even in the present era.
One can evaluate the thermodynamic pressure of neutrinos whether they are relativistic or non-relativistic in the present epoch by evaluating (\ref{Omeganu}) in the relativistic $E=p+m^2/(2p)$ or non-relativistic $E=m+p^2/(2m)$ limits.

Beacom et al \cite{Beacom}  have raised the possibility that if neutrinos  have a large interaction with some light boson then they do not decouple at $T\sim 1 {\rm MeV}$ but can continue to be at thermal equilibrium with the radiation bath of the bosons till the present epoch. This possibility   is not ruled out by big-bang-nucleosynthesis and CMB observations. The neutrino distribution function in this case would be $f(p)=(\exp(E/T) +1)^{-1}$  and the pressure of the thermally coupled neutrinos can be evaluated in the for the two possibilities that neutrinos are either relativistic or non-relativistic in the present epoch. We study the case of un-decoupled neutrinos as an interesting example of situation where the neutrino pressure is different compared to the decoupled neutrinos (although the decoupled neutrinos also have a relativistic-thermal distribution) .
In the following sections we give examples of $m(\phi)$ for the case of relativistic and non-relativistic neutrinos
which can give rise to a net negative pressure of the required magnitude.

\section{ Non-relativistic neutrinos}
We choose the neutrino mass function as
\be
m(\sigma)= -\frac{v^2}{\epsilon\, M} \, \log\left(\frac{\sigma M^2}{v^3}(1-\epsilon)\right)
\label{mnu-nr}
\ee
where  $M$ is some large mass scale, $v=246.2{\rm GeV}$ and $\epsilon$ is a small dimensionless parameter.

\subsection{Non-relativistic decoupled neutrinos}
\bea
\Omega_\nu(\phi)=-P_\nu &=& -g_i \int \frac{d^3 p}{(2 \pi)^3} \,\frac{1}{3}\left(\frac{p^2}{m}- \frac{p^4}{2m^3} \right) \frac{1}{\exp(p/T) +1}\nonumber\\
&=& - \frac{ 45\, \zeta(5)}{2 \pi^2 }\,\frac{T^5}{m}+ \frac{2835\, \zeta(7)}{8 \pi^2 }\, \frac{T^7}{m^3}
\eea
where we have taken the distribution function of neutrinos to be relativistic, and we have taken $g_i=6$ assuming that all three species of neutrinos are non-relativistic in the present epoch.

Then the total thermodynamic potential (to the leading order in $m/T$) is
\be
\Omega(\sigma)= \lambda v^2 \sigma^2 - \frac{ 45\, \zeta(5)}{2 \pi^2 }\,\frac{T^5}{m}
\label{Omega-nr}
\ee
has minima  at the value of $\sigma=\sigma_m$ given by the equation $\Omega^\prime=0$,
\be
2 \lambda v^2 \sigma_m = -\, \frac{ 45\, \zeta(5) \,T^5}{2 \pi^2 }\,  \frac{1}{ m(\sigma_m)\, \log\left(\frac{\sigma_m M^2}{v^3}(1-\epsilon)\right)}
%\left(\frac{1}{\frac{\sigma_m M^2}{v^3}- \epsilon}\right)\frac{ M^2}{v^3}
\label{Omegamin-nr}
\ee
This equation can be re-written in terms of the neutrino mass $m(\sigma_m)$ (equation (\ref{mnu-nr})) and the dark energy density
$\rho_\lambda$ (equation(\ref{rho}) ) as,
\be
m(\sigma_m) \, \log\left(\frac{\sigma_m M^2}{v^3}(1-\epsilon)\right)
%\left(\frac{\sigma_m M^2}{v^3}- \epsilon\right)\left(\frac{v^3}{\sigma_m M^2}\right)
 = -\frac{ 45\, \zeta(5) T^5}{2 \pi^2 }\, \,\frac{1}{2 \rho_\lambda}= - 1.0 \times 10^{-8} \, {\rm eV}
\ee
A solution of this  equation can be obtained  (in the approximation $\epsilon \ll 1$), by choosing
\be
\frac{\sigma_m M^2}{v^3}=1,
\label{sigmaM2-nr}
\ee
and using $\log(1+x) \simeq x$ for small $x$, to get
\be
m(\sigma_m) \,\epsilon=1.0 \times 10^{-8} \, {\rm eV}
\label{meps}
\ee
Writing (\ref{sigmaM2-nr}) in terms of the dark energy density (\ref{rho}) we have,
\be
\rho_\lambda= \frac{\lambda v^8}{M^4}
\label{rho-nr}
\ee
from (\ref{rho-nr}) we see that taking  $ M=3\times 10^{16} {\rm GeV}\lambda^{1/4}$ we get the correct magnitude $\rho_\lambda= (0.002{\rm eV})^4$ for the dark energy density. The neutrino mass (\ref{mnu-nr}) is
\be
m(\sigma_m)= \frac{v^2}{M}=0.002 {\rm eV} \left(\frac{3 \times 10^{16} {\rm GeV} \, \lambda^{1/4}}{M} \right)
\ee
The two observable quantities $\rho_\lambda$ and $m$ depend on only one parameter $M$. The other parameter
determines the temperature at which the $\Omega$ minimization takes place. If $m=0.002 {\rm eV}$ then from the minimization condition (\ref{Omegamin-nr}) and (\ref{meps}) we see that taking $\epsilon=0.5\times 10^{-5}$ ensures that the minima of the
thermodynamic potential occurs at the present era (when the neutrino temperature is $T=1.7\times 10^{-4} {\rm eV}$).

\subsection{Non-relativistic non-decoupled neutrinos}
If neutrinos have a large interaction with some light boson then they do not decouple at $T\sim 1 {\rm Mev}$ but can continue to be at thermal equilibrium with the radiation bath of the bosons till the present epoch. This possibility was studied by Beacom et al \cite{Beacom} who found that this possibility is not ruled out by big-bang-nucleosynthesis and CMB observations.  Assuming that the bosons-neutrino fluid was relativistic at some temperature $T=(4/11)^{1/3} T_\gamma$, and subsequently the neutrinos are non-relativistic then from entropy conservation it follows that that the present common temperature of the non-relativistic neutrinos and relativistic bosons is $T= (25/11)^{1/3}\,T_\gamma= 3.8 \times 10^{-4} \, {\rm eV}$ \cite{Beacom}.

If the neutrinos are non-relativistic at the present epoch then the neutrino pressure at thermal equilibrium is describe by
\be
-\Omega_\nu=P_\nu= 6 \left(\frac{m(\phi)}{2 \pi} \right)^{3/2}\,T^{5/2} \exp\left(\frac{-m(\phi)}{T} \right)
\ee
The minima of the thermodynamic potential occurs at $(\Omega_\phi +\Omega_\nu)^\prime=0$, and is given by the solution of
\be
2 \lambda v^2 \sigma_m-  \frac{6}{(2 \pi)^{3/2}} m(\sigma_m)^{1/2} T^{5/2}\left(\frac{3}{2} -\frac{m(\sigma_m)}{T} \right) \exp\left( \frac{-m(\sigma_m)}{T}\right)\, \left( \frac{-v^2}{\epsilon M \sigma_m} \right) =0
\label{nrtomegaprime}
\ee
We choose the large mass scale $M$  in the neutrino mass function (\ref{mnu-nr}) such that
\be
 \frac{\sigma_m M^2}{v^3}=1
 \ee
 which implies that $M=3.0 \times 10^{16} {\rm GeV} \lambda^{1/4}$. Assuming the parameter $\epsilon \ll1 $ the expression (\ref{mnu-nr}) then gives us the neutrino mass,
 \be
 m(\sigma_m) = \frac{v^2}{M}= \frac{1}{\lambda^{1/4}} \,2 \times 10^{-3} {\rm eV}
 \label{nrtnumass}
 \ee
 With this value of neutrino mass the ratio $m(\sigma_m)/T= 5.2 \lambda^{-1/4}$   which is consistent with our assumption that neutrinos are non-relativistic. One can solve thermodynamic potential minimization equation (\ref{nrtomegaprime}) for the remaining parameter $\epsilon$ to obtain $\epsilon= 5.9 \times 10^{-5} $, which ensures that the thermodynamic minima is achieved in the present epoch  (when the neutrino temperature is $T_{\nu }= (25/11)^{1/3}\,T_\gamma= 3.8 \times 10^{-4} \, {\rm eV}$).

\section{ Relativistic neutrinos} Assume that at-least one of the neutrino masses is smaller than the neutrino temperature
$ T =(4/11)^{1/3} T_0= 1.7 \times 10^{-4}$ eV. We consider two possibilities (a) neutrinos decouple at $T\simeq 1{\rm MeV} $ and  have a relativistic thermal distribution  (b) they are in thermal equilibrium with the radiation in the present epoch as studied in reference \cite{Beacom}. For both the relativistic neutrino species we consider the mass dependence on the standard model Higgs of the form,
\be
m(\sigma)=-\frac {v^2}{\epsilon M}\, \log\left(\frac{\sigma M^2}{\epsilon v^3}(1 -\epsilon^2) \right)
\label{mnu-r}
\ee
where again   $M$ is some large mass scale, $v=246.2{\rm GeV}$ and $\epsilon$ is a small dimensionless parameter.

\subsection{Relativistic decoupled neutrinos}

The thermodynamic potential of relativistic neutrinos in terms of neutrino
mass $m(\phi)$ is obtain from (\ref{Omeganu}) in the relativistic limit $m\leq T$ is
\bea
\Omega_\nu(\phi)=-P_\nu &=& -g_i \int \frac{d^3 p}{(2 \pi)^3} \,\frac{1}{3}\left(p- \frac{m^2}{2p} \right) \frac{1}{\exp(p/T) +1}\nonumber\\
&=&  - \frac{7 \pi^2}{360 } T^4 +\frac{ T^2}{72} m^2
\eea
where we have taken $g_i=2$ for  one relativistic neutrino species.

Then the total thermodynamic potential is
\be
\Omega(\sigma)= \lambda v^2 \sigma^2 + \frac{ T^2}{72} m(\sigma)^2- \frac{7 \pi^2}{360 } T^4
\label{Omega-r}
\ee
The minima of $\Omega$ occurs at the value of $\sigma=\sigma_m$ given by the equation $\Omega^\prime=0$,
\be
 \lambda v^2 \sigma_m  -  \frac{ T^2 }{72 \sigma_m} m(\sigma_m) \frac {v^2}{\epsilon M}\, =0
\label{dOmega-r}
\ee
We can re-write equation (\ref{dOmega-r}) in terms of the neutrino mass $m(\sigma_m)$ and  dark energy density (\ref{rho}) as,
\be
-\log\left(\frac{\sigma_m M^2}{\epsilon\, v^3}(1 -\epsilon^2) \right)= \frac{T^2 m(\sigma_m)^2}{72 \,\rho_\lambda}
\label{log-r}
\ee
Equation (\ref{log-r}) has the solution
\be
\left(\frac{\sigma_m M^2}{\epsilon\, v^3}\right)=1
\label{M-r-1}
\ee
and
\be
\epsilon  =  \left( \frac{T m(\sigma_m)}{\sqrt{72}  \,\sqrt{\rho_\lambda}} \right)
\label{epsilon-r}
\ee
The neutrino mass (\ref{mnu-r}) reduces to the form,
\be
m(\sigma_m)= \frac{v^2}{M} \epsilon
\label{mnu2-r}
\ee
 Substituting for $m(\sigma_m)$ in (\ref{epsilon-r}) we can solve for the mass parameter $M$,
\be
M=\frac{T\, v^2}{\sqrt{72\, \rho_\lambda}}= 3.0 \times 10^{14} {\rm GeV}
\label{M-r}
\ee
Substituting the expression (\ref{M-r}) for $M$ in equation (\ref{M-r-1}) we can solve for $\epsilon$ which turns out to be
\be
\epsilon= \frac{T^2}{72 \sqrt{\lambda\,  \rho_\lambda}}= \frac{1}{\sqrt{\lambda}}\, 1.0 \times 10^{-4}
\label{epsilon2-r}
\ee
Using (\ref{M-r}) and (\ref{epsilon2-r}) in the  the expression (\ref{mnu2-r}),  the neutrino mass turns out to be,
\be
m(\sigma_m)= 2.0\times 10^{-5} \frac{{\rm eV}}{\sqrt{\lambda}}
\ee
which is consistent with the assumption that atleast  one species  neutrinos is relativistic in the present era.

\subsection{Relativistic non-decoupled neutrinos }
Now we consider the case of neutrinos which are still in thermal equilibrium with some massless hidden sector radiation and which are still relativistic.
The thermodynamic potential of relativistic neutrinos in terms of neutrino
mass $m(\phi)$ is obtain from (\ref{Omeganu}) in the relativistic limit $m\leq T $ is 
\be
\Omega_\nu(\phi)= -g_i \int \frac{d^3 p}{(2 \pi)^3} \,\frac{1}{3}\left(\frac{p^2}{(p^2+m^2)^{1/2}}\right) \frac{1}{\exp((p^2 +m^2)^{1/2}/T) +1}
\ee
and expanding the integrand in powers of $m/p$.
Taking $g_i=2$ for  one relativistic neutrino species we obtain in the leading order in $m/T,$
\be
\Omega_\nu(\phi)= -P_\nu= - \frac{7 \pi^2}{360 } T^4 +\frac{ T^2}{24} m^2
\ee
Then the total thermodynamic potential is
\be
\Omega(\sigma)= \lambda v^2 \sigma^2 + \frac{ T^2}{24} m(\sigma)^2- \frac{7 \pi^2}{360 } T^4
\label{Omega-rt}
\ee
The minima of $\Omega$ occurs at the value of $\sigma=\sigma_m$ given by the equation $\Omega^\prime=0$,
\be
 \lambda v^2 \sigma_m  -  \frac{ T^2 }{24\, \sigma_m} m(\sigma_m) \frac {v^2}{\epsilon M}\, =0
\label{dOmega-rt}
\ee
We can re-write equation (\ref{dOmega-rt}) in terms of the neutrino mass $m(\sigma_m)$ and  dark energy density (\ref{rho}) as,
\be
-\log\left(\frac{\sigma_m M^2}{\epsilon\, v^3}(1 -\epsilon^2) \right)= \frac{T^2 m(\sigma_m)^2}{24 \,\rho_\lambda}
\label{log-rt}
\ee
Equation (\ref{log-rt}) has the solution
\be
\left(\frac{\sigma_m M^2}{\epsilon\, v^3}\right)=1
\label{M-rt-1}
\ee
and
\be
\epsilon  =  \left( \frac{T m(\sigma_m)}{\sqrt{24}  \,\sqrt{\rho_\lambda}} \right)
\label{epsilon-rt}
\ee
The neutrino mass (\ref{mnu-r}) is of the form,
\be
m(\sigma_m)= \frac{v^2}{M} \epsilon
\label{mnu2-rt}
\ee
Substituting (\ref{mnu2-rt}) for $m(\sigma_m)$ in (\ref{epsilon-rt}) we can solve for the mass parameter $M$,
\be
M=\frac{T\, v^2}{\sqrt{24\, \rho_\lambda}}= 5.2 \times 10^{14} {\rm GeV}
\label{M-rt}
\ee
Substituting the expression (\ref{M-rt}) for $M$ in equation (\ref{M-rt-1}) we can solve for $\epsilon$ which turns out to be
\be
\epsilon= \frac{T^2}{24\, \sqrt{\lambda\, \rho_\lambda}}= \frac{1}{\sqrt{\lambda}}\, 3.0 \times 10^{-4}
\label{epsilon2-rt}
\ee
Using (\ref{M-rt}) and (\ref{epsilon2-rt}) in  the expression (\ref{mnu2-rt}) we find that  the neutrino mass is
\be
m(\sigma_m)= 6.9\times 10^{-5} \frac{{\rm eV}}{\sqrt{\lambda}}
\ee
This is smaller than the neutrino temperature $T=1.7 10^{-4} {\rm eV}$ which is consistent with the assumption that the neutrino species is relativistic at the present epoch.

\section{Conclusions}
In this paper we have studied the possibility that observed the dark energy is due the shift of the Higgs field from the zero-minima of the Higgs potential due to the pressure of neutrinos. We have studied four different scenarios for the cosmological neutrino background. Neutrinos could be  relativistic and decoupled, non-relativistic and decoupled as in the standard cosmology. We have also considered the possibility that neutrinos are still in thermal equilibrium with some hidden sector radiation \cite{Beacom} for comparison. For our model to work the neutrino mass as function of the Higgs field must have a very steep gradient. We have considered logarithmic neutrino mass functions which meet this criterion. Interestingly the same mass model works for relativistic neutrinos whether they are decoupled or interacting. The same is true for the non-relativistic case, one neutrino mass model works for both the decoupled and the interacting neutrino scenarios even though the non-relativistic interacting neutrinos have a Boltzmann suppression in the number density. The neutrino mass function has a mass scale M, and dimensionless small parameter $\epsilon$ as free parameters. For the case of non-relativistic neutrinos $M = 3\times 10^{16} {\rm GeV} \lambda^{1/4}$  while for the case of relativistic neutrinos $M \simeq 10^{14} {\rm GeV}$. The GUT scale comes out naturally owing to the numerical relation $(\rho_\lambda)^{1/4} \simeq v^2/M$. Neutrino which is consistent with this scenario  turns out to be $m_\nu  =2\times 10^{-3}{\rm eV}$ (for the non-relativistic cases). This is in the mass range consistent with the solution of the atmospheric neutrino problem.The parameter $\epsilon \sim 10^{-5} -10^{-4}$ is put in to ensure that the minima-of the Higgs-neutrino thermodynamic potential is achieved at the present era.

In this paper we explain the dark energy in terms of known particles. The neutrino mass must be a steep function of the Higgs field  in order achieve this. It would be interesting to see if there are well motivated  particle physics theory which gives rise to such neutrino mass functions.

\end{document}